\documentclass[twocolumn]{aastex631}

\usepackage{url}

\usepackage{amsmath}
\usepackage{float}
\usepackage{graphicx}
\usepackage{multirow}
\usepackage{soul}
\usepackage{rotating}
\usepackage{hyperref}

\newcommand{\feh}{\ensuremath{\left[{\rm Fe}/{\rm H}\right]}}

\newcommand{\teff}{\ensuremath{T_{\rm eff}}}

\newcommand{\logg}{\ensuremath{\log g_*}}

\newcommand{\mj}{\ensuremath{\,M_{\rm J}}}
\newcommand{\mearth}{\ensuremath{\,M_{\Earth}}}

\newcommand{\mplanet}{\ensuremath{\,M_{\rm P}}}

\newcommand{\cosi}{\ensuremath{\cos{i}}}

\newcommand{\kepler}{{\it Kepler}}
\newcommand{\ktwo}{{\it K2}}
\newcommand{\tess}{{\it TESS}}

\newcommand{\mstar}{\ensuremath{M_{*}}}
\newcommand{\rstar}{\ensuremath{R_{*}}}
\newcommand{\ar}{\ensuremath{a/R_*}}
\newcommand{\vsini}{\ensuremath{v\sin{i_*}}}
\newcommand{\exofasttwo}{{\tt EXOFASTv2}}

\newcommand{\degrees}{\ensuremath{^{\circ}}}

\newcommand{\Nall}{78}
\newcommand{\Nhj}{55}
\newcommand{\Nwj}{23}
\newcommand{\Nhjhs}{36}
\newcommand{\Nwjhs}{9}

\newcommand{\NhjhsAligned}{18}

\newcommand{\SigHotStar}{3.4}

\begin{document}
\title{Single-Star Warm-Jupiter Systems Tend to Be Aligned, Even Around Hot Stellar Hosts:\\ No \teff\,-$\lambda$ Dependency\footnote{This paper includes data gathered with the 6.5 meter Magellan Telescopes located at Las Campanas Observatory, Chile.}}

\author[0000-0002-0376-6365]{Xian-Yu Wang} 
\affiliation{Department of Astronomy, Indiana University, 727 East 3rd Street, Bloomington, IN 47405-7105, USA}

\author[0000-0002-7670-670X]{Malena Rice}
\affiliation{Department of Astronomy, Yale University, 219 Prospect Street, New Haven, CT 06511, USA}

\author[0000-0002-7846-6981]{Songhu Wang}
\affiliation{Department of Astronomy, Indiana University, 727 East 3rd Street, Bloomington, IN 47405-7105, USA}

\author[0000-0001-8401-4300]{Shubham Kanodia}
\affiliation{Carnegie Institution for Science, Earth \& Planets Laboratory, 5241 Broad Branch Road NW, Washington, DC 20015, USA}

\author[0000-0002-8958-0683]{Fei Dai}  
\affiliation{Institute for Astronomy, University of Hawai`i, 2680 Woodlawn Drive, Honolulu, Hawaii 96822, USA}

\author[0000-0002-9632-9382]{Sarah E. Logsdon}
\affiliation{NSF's National Optical-Infrared Astronomy Research Laboratory, 950 N. Cherry Ave., Tucson, AZ 85719, USA}

\author[0000-0001-9580-4869]{Heidi Schweiker}
\affiliation{NSF's National Optical-Infrared Astronomy Research Laboratory, 950 N. Cherry Ave., Tucson, AZ 85719, USA}

\author[0009-0008-2801-5040]{Johanna K. Teske}
\affiliation{Carnegie Institution for Science, Earth \& Planets Laboratory, 5241 Broad Branch Road NW, Washington, DC 20015, USA}
\affiliation{The Observatories of the Carnegie Institution for Science, 813 Santa Barbara Street, Pasadena, CA 91101, USA}

\author[0000-0003-1305-3761]{R. Paul Butler}
\affiliation{Carnegie Institution for Science, Earth \& Planets Laboratory, 5241 Broad Branch Road NW, Washington, DC 20015, USA}

\author[0000-0002-5226-787X]{Jeffrey D. Crane}
\affiliation{The Observatories of the Carnegie Institution for Science, 813 Santa Barbara Street, Pasadena, CA 91101, USA}

\author[0000-0002-8681-6136]{Stephen Shectman}
\affiliation{The Observatories of the Carnegie Institution for Science, 813 Santa Barbara Street, Pasadena, CA 91101, USA}

\author[0000-0002-8964-8377]{Samuel N. Quinn}
\affiliation{Center for Astrophysics ${\rm \mid}$ Harvard {\rm \&} Smithsonian, 60 Garden Street, Cambridge, MA 02138, USA}

\author{Veselin Kostov} 
\affiliation{NASA Goddard Space Flight Center, 8800 Greenbelt Rd, Greenbelt, MD 20771, USA}
\affiliation{SETI Institute, 189 Bernardo Ave, Suite 200, Mountain View, CA 94043, USA}

\author[0000-0002-4047-4724]{Hugh~P.~Osborn}
\affiliation{Department of Physics and Kavli Institute for Astrophysics and Space Research, Massachusetts Institute of Technology, Cambridge, MA 02139, USA}
\affiliation{Physics Institute, Universität Bern, Gesellschaftsstrasse 6, 3012 Bern, Switzerland}

\author[0000-0003-1748-5975]{Robert~F.~Goeke}
\affiliation{Department of Physics and Kavli Institute for Astrophysics and Space Research, Massachusetts Institute of Technology, Cambridge, MA 02139, USA}

\author[0000-0003-3773-5142]{Jason~D.~Eastman}
\affiliation{Center for Astrophysics ${\rm \mid}$ Harvard {\rm \&} Smithsonian, 60 Garden Street, Cambridge, MA 02138, USA}

\author[0000-0002-1836-3120]{Avi~Shporer}
\affiliation{Department of Physics and Kavli Institute for Astrophysics and Space Research, Massachusetts Institute of Technology, Cambridge, MA 02139, USA}

\author[0000-0003-2196-6675]{David Rapetti}
\affiliation{NASA Ames Research Center, Moffett Field, CA 94035, USA}
\affiliation{Research Institute for Advanced Computer Science, Universities Space Research Association, Washington, DC 20024, USA}

\author[0000-0001-6588-9574]{Karen A.\ Collins}
\affiliation{Center for Astrophysics ${\rm \mid}$ Harvard {\rm \&} Smithsonian, 60 Garden Street, Cambridge, MA 02138, USA}

\author[0000-0001-8621-6731]{Cristilyn N.\ Watkins}
\affiliation{Center for Astrophysics \textbar \ Harvard \& Smithsonian, 60 Garden Street, Cambridge, MA 02138, USA}

\author[0009-0009-5132-9520]{Howard M. Relles}
\affiliation{Center for Astrophysics \textbar \ Harvard \& Smithsonian, 60 Garden Street, Cambridge, MA 02138, USA}

\author[0000-0003-2058-6662]{George~R.~Ricker} 
\affiliation{Department of Physics and Kavli Institute for Astrophysics and Space Research, Massachusetts Institute of Technology, Cambridge, MA 02139, USA}

\author[0000-0002-6892-6948]{Sara~Seager} 
\affiliation{Department of Physics and Kavli Institute for Astrophysics and Space Research, Massachusetts Institute of Technology, Cambridge, MA 02139, USA}
\affiliation{Department of Earth, Atmospheric and Planetary Sciences, Massachusetts Institute of Technology, Cambridge, MA 02139, USA}
\affiliation{Department of Aeronautics and Astronautics, MIT, 77 Massachusetts Avenue, Cambridge, MA 02139, USA}

\author[0000-0002-4265-047X]{Joshua~N.~Winn} 
\affiliation{Department of Astrophysical Sciences, Princeton University, 4 Ivy Lane, Princeton, NJ 08544, USA}

\author[0000-0002-4715-9460]{Jon~M.~Jenkins} 
\affiliation{NASA Ames Research Center, Moffett Field, CA 94035, USA}

\correspondingauthor{Songhu Wang}
\email{sw121@iu.edu}

\begin{abstract}
The stellar obliquity distribution of warm-Jupiter systems is crucial for constraining the dynamical history of Jovian exoplanets, as the warm Jupiters' tidal detachment likely preserves their primordial obliquity. However, the sample size of warm-Jupiter systems with measured stellar obliquities has historically been limited compared to that of hot Jupiters, particularly in hot-star systems. In this work, we present newly obtained sky-projected stellar obliquity measurements for warm-Jupiter systems, TOI-559, TOI-2025, TOI-2031, TOI-2485, TOI-2524, and TOI-3972, derived from the Rossiter–McLaughlin effect, and show that all six systems display alignment with a median measurement uncertainty of 13\degrees. {Combining these new measurements with the set of previously reported stellar obliquity measurements, our analysis reveals that single-star warm-Jupiter systems tend to be aligned, even around hot stellar hosts. This alignment exhibits a 3.4-$\sigma$ deviation from the $\teff-\lambda$ dependency observed in hot-Jupiter systems, where planets around cool stars tend to be aligned, while those orbiting hot stars show considerable misalignment.} The current distribution of spin-orbit measurements for Jovian exoplanets indicates that misalignments are neither universal nor primordial phenomena affecting all types of planets. The absence of misalignments in single-star warm-Jupiter systems further implies that many hot Jupiters, by contrast, {have experienced a dynamically violent history}.
\end{abstract}


\section{Introduction} \label{sec:intro}
\begin{deluxetable*}{lcccccccccc}
\tabletypesize{\scriptsize}
\tablewidth{0pt}
\tablecaption{Log of Observations}
\startdata
\\
\multicolumn{10}{c}{Spectroscopy Observations$^{a}$}\\
\hline
Name       &     Facility/Instrument&           Time &       N$^{b}$    &  Exposure time &Seeing& Airmass& Moon Phase & Moon Distance  &  S/N$^{c}$  \\
       &     &         (UT)     &        &  (s) & ($''$) & & (\%) & ($\degrees$)  &  \\
\hline
\multicolumn{10}{c}{In-Transit}\\
TOI-559       &     Magellan II/PFS&       2023-09-26 03:16:41 - 09:10:21 &        15     &  1520 &1.4-2.0    & 1.00-1.74& 79 & 76  &  46  \\
TOI-2025      &     WIYN/NEID&             2023-06-14 04:42:20 - 11:04:24 &        20     & 1123  &1.2-2.3   & 1.62-1.75& 15 & 79 & 20 \\
TOI-2031      &     WIYN/NEID&             2023-10-17 01:46:26 - 07:41:43 &        16     &  1290 &0.7-1.4    & 1.54-1.70& 3 & 76  &  23 \\
TOI-2485      &     WIYN/NEID&             2023-04-06 03:00:02 - 10:50:09 &        16     &  1800 &0.3-2.0    & 1.01-2.69& 99 & 31  &  50 \\
TOI-2524      &     WIYN/NEID&             2024-03-20 02:32:06 - 08:32:32 &        18     & 1200  &0.8-1.6    & 1.21-2.24  & 79 & 41 & 11 \\
TOI-3972      &     WIYN/NEID&             2022-12-21 03:15:02 - 08:06:24 &        18     &  1000 &0.6-1.1    & 1.16-2.18& 13 & 65  &  27 \\
TOI-2025      &     NOT/FIES&              2021-08-08 22:06:23 - 03:37:09$^{d}$ &        16     &\multicolumn{4}{c}{\cite{Knudstrup2022}} \\
\\
\multicolumn{10}{c}{Out-of-Transit}\\
TOI-559       &     FLWO/TRES&           2019-01-28 - 2019-09-12 &      3    &\multicolumn{4}{c}{\cite{Ikwut2022}}   \\
TOI-559                               &     CTIO/CHIRON&           2019-01-27 - 2019-09-20 &      22   &\multicolumn{4}{c}{\cite{Ikwut2022}}   \\
TOI-2025       &     FLWO/TRES&        2020-07-30 - 2021-04-13 &      16    &\multicolumn{4}{c}{\cite{Rodriguez2023}}  \\
TOI-2025       &     NOT/FIES&        2020-10-01 - 2022-06-02 &      39    &\multicolumn{4}{c}{\cite{Knudstrup2022}}   \\
TOI-2025       &     WIYN/NEID&        2023-04-04 - 2023-04-04 &      6    &  \\
TOI-2524       &     CTIO/CHIRON&      2020-12-26 - 2021-03-09 &      8    &\multicolumn{4}{c}{\cite{Eberhardt2023}}   \\
TOI-3972       &     WIYN/NEID&        2023-10-18 - 2024-01-29 &      13    &   && &  &  &   \\
\hline
\\
\multicolumn{10}{c}{TESS Observations}\\
\hline
Name       &     \multicolumn{9}{c}{Sector (source, cadence in seconds)} \\
\hline
TOI-559       &    \multicolumn{2}{l}{04 (TESS-SPOC, 1800); 31 (SPOC, 120)}  \\
TOI-2025    &     \multicolumn{9}{l}{14 (TESS-SPOC, 1800); 18 (TESS-SPOC, 1800); 19 (TESS-SPOC, 1800); 20 (TESS-SPOC, 1800); 24 (TESS-SPOC, 1800);} \\
TOI-2025    &     \multicolumn{9}{l}{25 (TESS-SPOC, 1800); 26 (TESS-SPOC, 1800);  40 (SPOC, 120);       47 (SPOC, 120);       52 (SPOC, 120); 53 (SPOC, 120); } \\
TOI-2025    &     \multicolumn{9}{l}{58 (SPOC, 120); 59 (SPOC, 120); 60 (SPOC, 120); 74 (SPOC, 20)} \\
TOI-2031    &     \multicolumn{9}{l}{18 (TESS-SPOC, 1800); 19 (TESS-SPOC, 1800); 24 (TESS-SPOC, 1800); 25 (TESS-SPOC, 1800); 26 (TESS-SPOC, 1800);} \\
TOI-2031   & \multicolumn{9}{l}{52 (SPOC, 120); 53 (SPOC, 120); 58 (SPOC, 120); 59 (SPOC, 120); 60 (SPOC, 120); 73 (SPOC, 120)}\\
TOI-2485      &    \multicolumn{9}{l}{23 (TESS-SPOC, 1800); 50 (TESS-SPOC, 600)} \\
TOI-2524      &    \multicolumn{9}{l}{09 (QLP, 1800); 35 (QLP, 600); 45 (SPOC, 120); 46 (SPOC, 120); 62 (SPOC, 120); 72 (SPOC, 120) } \\
TOI-3972      &     \multicolumn{9}{l}{17 (TESS-SPOC, 1800); 18 (TESS-SPOC, 1800); 24 (TESS-SPOC, 600); 58 (SPOC, 120)} \\
\hline
\\
\multicolumn{10}{c}{Ground-Based Photometric Follow-Up Observations}\\
\hline
Name       &     Facility/Instrument&           Aperture  &Filter & Date &       Exposure time   & \multicolumn{2}{c}{Source}  \\
        &     &           (m)  &  & (UT) &  (s)         \\
\hline
TOI-559       &   PEST & 0.3048 & Rc & 2019-09-27 & 60 &\multicolumn{4}{c}{\cite{Ikwut2022}}\\
TOI-559       &   LCOGT SSO & 1.0 & Sloan z$\arcmin$ & 2019-10-18 & 35  &\multicolumn{4}{c}{\cite{Ikwut2022}}\\
TOI-559       &   LCOGT SSO & 1.0 & Sloan i$\arcmin$ & 2020-08-20 & 25  &\multicolumn{4}{c}{\cite{Ikwut2022}}\\
TOI-559       &   LCOGT SSO & 1.0 & Sloan i$\arcmin$ & 2020-08-27 & 25  &\multicolumn{4}{c}{\cite{Ikwut2022}}\\
TOI-2025& Kotizarovci	&	0.3	&   TESS &  2020-06-26	&30	&\multicolumn{4}{c}{\cite{Knudstrup2022,Rodriguez2023}} \\
TOI-2025& LCOGT TFN	    &  	0.4	&   g$\arcmin$	&2020-06-26	&60	&\multicolumn{4}{c}{\cite{Knudstrup2022,Rodriguez2023}} \\
TOI-2025& FLWO/KeplerCam	    &  	1.2	&   B	&2021-05-12 &20		&\multicolumn{4}{c}{\cite{Knudstrup2022,Rodriguez2023}} \\
TOI-2025& FLWO/KeplerCam	    &  	1.2	&   i$\arcmin$	&2021-05-12 &7		&\multicolumn{4}{c}{\cite{Knudstrup2022,Rodriguez2023}} \\
TOI-2025& GMU	    &  	0.8	&   R	&2021-05-12 &50		&\multicolumn{4}{c}{\cite{Knudstrup2022,Rodriguez2023}} \\
TOI-2025& CRCAO	    &  	0.61	&   R	&2021-05-12 &120		&\multicolumn{4}{c}{\cite{Knudstrup2022,Rodriguez2023}} \\
TOI-2025& CPO	    &  	0.61	&   V	&2021-09-18 &30		&\multicolumn{4}{c}{\cite{Knudstrup2022,Rodriguez2023}} \\
TOI-2025 &   LCOGT McD & 0.35 &  zs & 2023-06-14 & 55 &\multicolumn{4}{c}{This work}\\
TOI-2524& El Sauce	    &  	0.36	&  $R_c$	&2021-03-31		&180&\multicolumn{4}{c}{\cite{Eberhardt2023}} \\
TOI-2031 &   LCOGT McD & 0.35 &  ip & 2023-07-05 & 70 &\multicolumn{4}{c}{This work}\\
\hline
\enddata
\tablenotetext{a}{For already-published data, we have added the corresponding reference at the end of each data row.}
\tablenotetext{b}{The number of exposures.}
\tablenotetext{c}{Typical signal-to-noise (S/N) at 5530 $\rm \AA$ for NEID and in the iodine region (5010 - 6010 $\rm \AA$) for PFS.}
\tablenotetext{d}{This observation started on UT 2021-08-08 at 22:06:23 and ended on UT 2021-08-09 at 03:37:09. }
\label{obslog}
\end{deluxetable*}
The Solar System planets lie on near-coplanar orbits that are close to alignment (within $\sim6^{\circ}$) with the Sun's equator, offering a seemingly coherent framework for planetary formation \citep{Kant1755, Laplace1796}. In contrast, the landscape of exoplanetary systems is remarkably varied, showing a puzzling range of spin-orbit orientations for hot Jupiters (see \citealt{WinnFabrycky2015, Triaud2018, Albrecht2022} and references within). The origins of these orbital misalignments have sparked a contentious debate: are misalignments a unique feature of hot-Jupiter systems due to their specific formation processes and dynamical histories, including planet-planet scattering \citep{Rasio1996, Beauge2012}, Kozai-Lidov oscillations \citep{Wu2003, Fabrycky2007, Naoz2016}, and secular interactions \citep{WuLithwick2011, Petrovich2015}? Or, do they instead hint at more ubiquitous processes, such as chaotic accretion \citep{Bate2010, Thies2011, Fielding2015, Bate2018}, magnetic warping \citep{Foucart2011, Lai2011, Romanova2013, Romanova2021}, tilting by a companion star \citep{Borderies1984, Lubow2000, Batygin2012, Matsakos2017}, or internal-gravity-wave-induced tumbling \citep{Lai2012, Lin2017, Damiani2018}, that may affect a wide array of planetary systems?

The relationship between the spin-orbit orientation of a system and the stellar host's properties adds another layer to the mystery. Previous studies have demonstrated that hot Jupiters around hot stars,  which possess thin convective envelopes and experience rapid rotation, span a wide range of spin-orbit angles, while those around cool stars tend to be aligned \citep{Schlaufman2010, Winn2010}. Could this trend indicate that tidal interactions have played a significant role in realigning the cool-star systems {\citep{Albrecht2012, RogersANDLin2013, Xue2014, Valsecchi2014, Li2016, Anderson2021, Wang2021, Rice2022, Spalding2022, zanazzi2024damping}?} Or, could this trend arise due to hot-star systems being dynamically ``hotter'' -- that is, could it be that these systems are more likely to form multi-giant planet systems, subsequently leading to inclination excitation from post-disk dynamical interactions \citep{Wu2023HJsNotAlone}? 

Warm Jupiters, defined in this work as wide-orbiting ($11\leq a/R_*  \leq200$) Jovian-mass ($0.3M_J \leq \mplanet \leq 13M_J$) planets\footnote{The $0.3 \, M_J$ or 100 \mearth\, minimum mass limit was adopted following \cite{Helled2023}, and is further motivated by the identified transition region around 95 - 150 \mearth\, from a series of statistical analyses (e.g., \citealt{Weiss2014MRrelation, Hatzes2015, Chen2017, Bashi2017, Muller2024}). The upper mass limit is set to 13 \mj, the minimum mass required for a brown dwarf to ignite deuterium fusion. We use the $\ar = 11$ division between hot and warm Jupiters throughout this work as a cutoff that has been empirically identified, following \citet{Rice2022WJs_Aligned}, and that concurs with physically motivated limits for efficient tidal realignment (see e.g. Figure 11 of \citealt{zanazzi2024damping}).}, offer a compelling lens through which to scrutinize these questions. The comparatively wider orbital separations of warm Jupiters make them less prone to tidal realignment than hot Jupiters -- which fall within the same mass range, but lie within $a/R_*<11$ -- such that they do not suffer from the same tidal realignment degeneracy that plagues interpretations of aligned hot Jupiters \citep{Rice2021K2140}. {Additionally,} warm Jupiters seem less likely to have undergone high-eccentricity migration --- a leading theory for hot Jupiter migration (see \citealt{Dawson2018} and references within) --- making them even more valuable for teasing apart the origins and evolution of spin-orbit misalignment.

Historically, the scarcity of spin-orbit measurements for warm Jupiters has stemmed from observational challenges, including the day/night cycle and weather limitations of ground-based transit surveys (e.g., SuperWASP, \citealt{Pollacco2006}; HATNet, \citealt{Bakos2004}; HATSouth, \citealt{Bakos2013}; TrES, \citealt{Alonso2004}; KELT, \citealt{Pepper2007}; XO, \citealt{McCullough2005} and, CSTAR, \citealt{Wang2014}) and the $\kepler$ mission's ``long stare'' observing strategy that identified transiting planets primarily around fainter host stars \citep{Borucki2010Sci}, which are not conducive to precise follow-up studies of wide-orbiting planets with long transits. However, this landscape has been transformed by the advent of the $\ktwo$ \citep{Howell2014} and $\tess\,$ \citep{Ricker2015} missions, which have successfully identified warm Jupiters around brighter host stars across the sky (e.g., K2-99 b, \citealt{Smith2017}; K2-114 b and K2-115 b, \citealt{Shporer2017}; K2-140 b, \citealt{Giles2018}; K2-232 b, \citealt{Brahm2018}; TOI-481 b and TOI-892 b, \citealt{Brahm2020}; TOI-558 b and TOI-559 b, \citealt{Ikwut2022}, TOI-1478 b, \citealt{Rodriguez2021FiveNewHotGiant}, and 55 warm Jupiter candidates listed in \citealt{Dong2021}).

In this context, the Stellar Obliquities in Long-period Exoplanet Systems (SOLES) survey \citep{Rice2021K2140, WangX2022WASP148, Rice2022WJs_Aligned, Rice2023Q6, Hixenbaugh2023, Dong2023, Wright2023, Rice2023TOI2202, Lubin2023, Hu2024PFS, radzom2024evidence, ferreira2024soles} was initiated to broaden the sample of Rossiter–McLaughlin (RM) measurements, especially for planets with wide orbital separations. This paper represents the most recent contributions from the SOLES survey, adding six new spin-orbit angle measurements for warm Jupiters (TOI-559 b, TOI-2025 b, TOI-2031 b, TOI-2485 b, TOI-2524 b, and TOI-3972 b\footnote{The Jovian natures of TOI-2031 b, TOI-2485 b, and TOI-3972 b have been confirmed in \cite{toi2031}, \cite{toi2485}, and \cite{toi3972}, respectively.}) that orbit hosts spanning a range of stellar types. Our findings reinforce the recent evidence that warm Jupiters in single-star systems are preferentially aligned relative to hot Jupiters, as first demonstrated by \cite{Rice2022WJs_Aligned}. 
We crucially extend this observation to demonstrate the same trend for warm Jupiters around hot stars, showing a {3.4}-$\sigma$ confidence difference with the \teff-$\lambda$ dependency observed in hot-Jupiter systems.

\section{Observations} \label{sec:obs}

\subsection{Spectroscopy Observation}\label{sec:Transit_Spectroscopy_Observation}

\subsubsection{PFS observation}
The transit spectroscopy sequence across the transit of TOI-559 b was conducted during UT September 26, 2023, using the Carnegie Planet Finder Spectrograph (PFS, \citealt{crane2006carnegie, crane2008carnegie, crane2010carnegie}) on the 6.5 m Magellan Clay telescope at Las Campanas Observatory (LCO), Chile. PFS, which has been on-sky since 2010, covers wavelengths from 391 to 734 nm with a default resolving power of $\sim$127,000 (1$\times$2 binning and 0.3$\arcsec$ slit). In our case, due to the faint magnitude of the star, we used $3\times3$ binning, resulting in a resolving power of $\sim$110,000. Our observing sequence provided a total of 15 RVs each with an exposure time of 1520 seconds, obtained with a moon phase of 79\% and a $76^{\circ}$ separation between the moon and TOI-559. Seeing ranged from $1.4-2\arcsec$. We also obtained an iodine-free template observation with PFS, which consisted of three 1000-second exposures during UT November 11, 2023 with seeing $0.55\arcsec$.

The spectral data reduction and radial velocity (RV) extraction for PFS spectra were performed using a customized pipeline described in \cite{butler1996attaining}. 

\begin{figure*}
    \centering
    \includegraphics[width=1\linewidth]{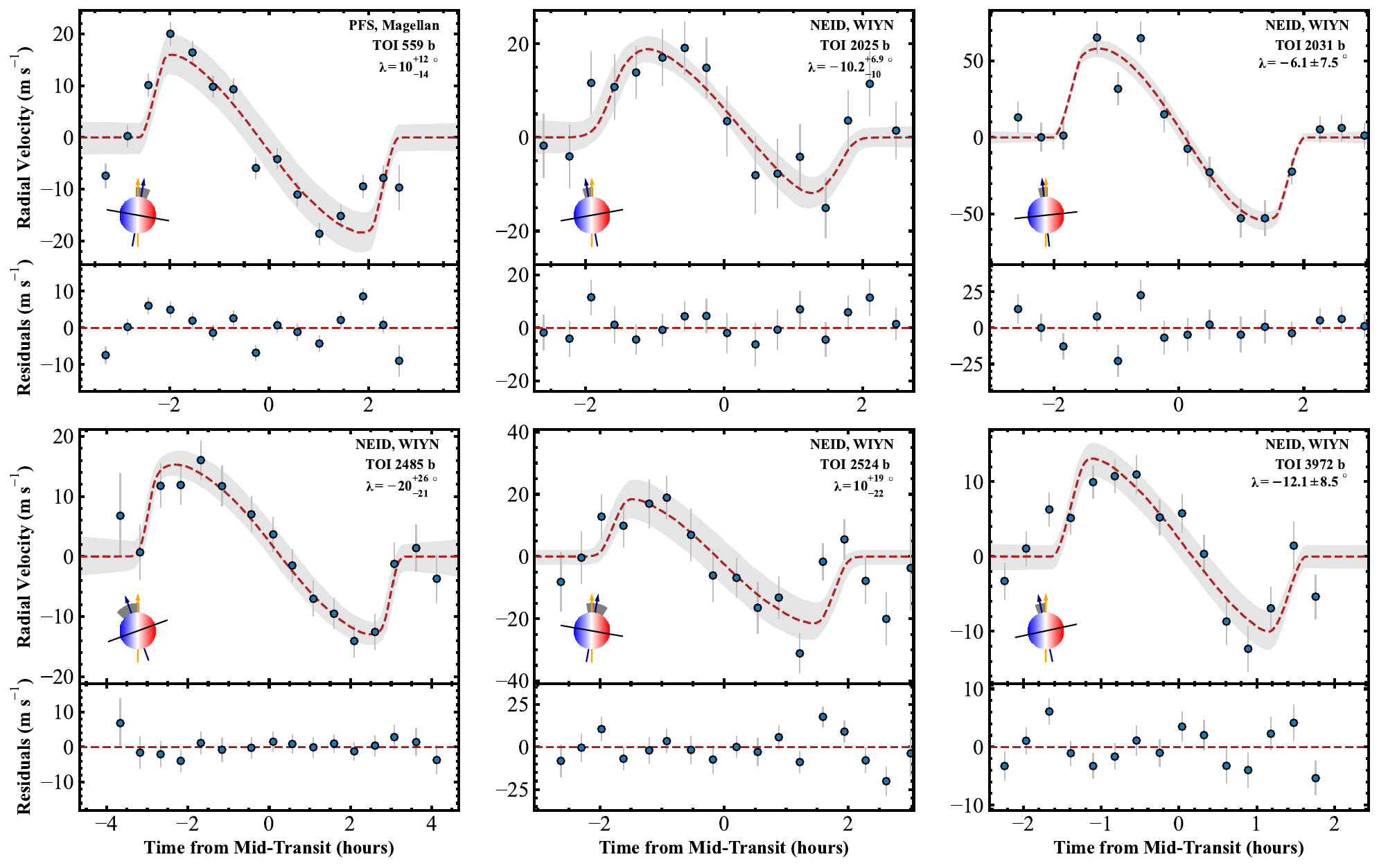}
    \caption{Rossiter-McLaughlin measurements for TOI-559 b, TOI-2025 b, TOI-2031 b, TOI-2485 b, TOI-2524 b and TOI-3972 b are shown as a function of hours from mid-transit. The RM data, with the Keplerian signal subtracted, are represented as blue dots. The median RM models are each depicted using red dashed lines. The 1$\sigma$ credible intervals are indicated by grey areas. In the lower-left corner, a schematic diagram illustrates the sky-projected stellar obliquity. Orange arrows indicate the projected stellar spin axis, while black arrows represent the planetary orbital axis, with its uncertainty shown in grey. Black lines illustrate the planetary orbital plane. Residuals are displayed beneath each panel. The RV data used in this work are available via this \href{https://github.com/wangxianyu7/Data_and_code/tree/main/WarmJupitersAreAligned}{link}.}
    \label{fig:RMs}
\end{figure*}

\subsubsection{NEID observation}
The transit spectroscopy observations of TOI-2025 b, TOI-2031 b, TOI-2485 b, TOI-2524 b and TOI-3972 b were conducted using the High-Resolution mode ($R \sim 110,000$) of the NEID spectrograph \citep{Schwab2016, Halverson2016} on the WIYN 3.5m telescope at Kitt Peak National Observatory in Arizona, USA. The fiber-fed \citep{Kanodia2018, Kanodia2023} and ultra-stable \citep{Stefannson2016, Robertson2019} NEID spectrograph has been operational since 2022, covering a wavelength range of 380 to 930 nm. Observations of these five targets were conducted on the following dates: UT June 14, 2023; October 17, 2023; April 6, 2023; March 20, 2024; and December 21, 2023, respectively, yielding a total of 88 measurements. Apart from the in-transit RVs, for TOI-3972, we collected 13 additional out-of-transit RVs from UT October 18, 2023 to January 29, 2024 using NEID. We also collected an additional six RVs for TOI-2025 on April 4, 2023, using NEID.

The NEID spectra were analyzed using version 1.3.0 of the NEID Data Reduction Pipeline (\texttt{NEID-DRP})\footnote{Detailed information available at \url{https://neid.ipac.caltech.edu/docs/NEID-DRP/}.}. The \texttt{DRP} adopted the cross-correlation function (CCF) method to derive the radial velocities. Barycentric-corrected radial velocities for re-weighted orders (\texttt{CCFRVMOD}) were extracted from the NExScI NEID Archive\footnote{Available at \url{https://neid.ipac.caltech.edu/}.}. 

A comprehensive summary of the observations conducted by PFS and NEID is presented in Table~\ref{obslog}. The resulting RVs, available through the Data behind the Figure program (see caption of Figure~\ref{fig:RMs}), are displayed in Figure~\ref{fig:RMs}.

\section{Photometry}

The \tess\, data used in this analysis were reduced by the \tess\, Science Processing Operations Center (SPOC;
\citealt{Jenkins2016}) at NASA Ames Research Center to obtain Presearch Data Conditioning simple aperture photometry (PDCSAP; \citealt{Smith2012, Stumpe2012, Stumpe2014}) light curves. Table~\ref{obslog} indicates the cadence at which data were acquired for each TOI and for each sector. The SPOC light curves were obtained at 2-min cadence and the TESS-SPOC light curves from Full Frame Images (FFIs; \citealt{Caldwell2016}). The transit signatures of TOIs 559 b, 2031 b, 2485 b, and 3972 b were all initially detected by the \tess\, Quick Look Pipeline (QLP; \citealt{Huang2020QLP,Huang2020QLP2}) and alerted by the \tess\, project to the community \citep{Guerrero2021}. Note that \tess\, Year 2 sectors (14-26) are affected by a sky background bias which starting in Sector 27 was addressed by the algorithm documented in the associated \tess\, Data Release Note DR38\footnote{\url{ https://archive.stsci.edu/missions/tess/doc/tess_drn/tess_sector_27_drn38_v02.pdf}}. We thus analyzed the target pixel and light curve files of the affected targets and sectors (TOI-2031, Sectors 18, 19, 24, 25, 26; TOI-3972, Sectors 17, 18, 24; and TOI-2485, Sector 23) to determine whether there was a significant effect in these cases. We found that the bias induced on the planetary radii for TOI-2031 ($<$1$\%$) and TOI-2485 ($\sim$0.3$\%$) was small compared to the respective overall errors ($\sim$3.2$\%$ and $\sim$5$\%$; see Table 2). However, for TOI-3972 since this bias was higher, we accounted for it in the fits, reducing the overall error on the planetary radius from $\sim$3.4$\%$ to $\sim$2.4$\%$ (see Table 2 for the latter). All \tess\ data used in this paper can be found in MAST: \href{https://archive.stsci.edu/doi/resolve/resolve.html?doi=10.17909/r8dh-5j34}{10.17909/r8dh-5j34}.

In this work, we also incorporated two new sets of ground-based photometry for TOI-2025 b and TOI-2031 b. Details on the observations and data reduction are provided in Appendix \ref{appendix:gb_photometry}.

\section{Stellar parameters}\label{sec:stellar}

\subsection{Synthetic spectral fitting by \texttt{iSpec}}

\par Co-added NEID spectra and iodine-free PFS spectra were adopted to determine the star's spectroscopic parameters, including stellar effective temperature  (\teff), surface gravity (\logg), metallicity (\feh), and projected rotational velocity (\vsini). We used the synthetic spectral fitting technique provided by the Python package  \texttt{iSpec} \citep{Blanco2014, Blanco2019} to measure these parameters.

We employed the SPECTRUM radiative transfer code \citep{Gray1994}, the MARCS atmosphere model \citep{gustafsson2008_MARCS}, and the sixth version of the GES atomic line list \citep{Heiter2021_GES}, all incorporated within \texttt{iSpec}, to create a synthetic model for the iodine-imprinted PFS spectra and the co-added NEID spectra (S/N: TOI-559, 188; TOI-2025, 83; TOI-2031, 88; TOI-2485, 151; TOI-2524, 41; TOI-3972, 96). We treated micro-turbulent velocities as a variable in our fitting process, allowing us to accurately represent the small-scale turbulent motions in the stellar atmosphere. Macro-turbulent velocities were determined using an empirical relationship that leverages established correlations with various stellar attributes \citep{Doyle2014Vmac}. We selected specific spectral regions to streamline the fitting process, focusing on the wings of the H$\alpha$, H$\beta$, and Mg I triplet lines, which are tracers of $\teff$ and $\logg$, as well as Fe I and Fe II lines that are crucial for constraining $\feh$ and $\vsini$. Spectroscopic parameters were refined using the Levenberg-Marquardt nonlinear least-squares fitting algorithm \citep{Markwardt2009}, which iteratively minimizes the $\chi^2$ value between the synthetic and observed spectra. The final spectroscopic parameters are detailed in Table~\ref{tab:results}.

\subsection{SED+MIST fit by \texttt{EXOFASTv2}}

To derive additional stellar parameters, such as stellar mass (\mstar) and radius (\rstar), we utilized the MESA Isochrones \& Stellar Tracks (MIST) model \citep{Choi2016mist, Dotter2016mist} in combination with a spectral energy distribution (SED) fitting approach. Photometry was compiled from various catalogs, including 2MASS \citep{Cutri2003}, WISE \citep{Cutri2014AllWISE}, \tess\, \citep{Ricker2015}, and Gaia DR3 \citep{GaiaCollaboration2023}. Gaussian priors based on our synthetic spectral fitting were applied to \teff\, and \feh, along with the parallax from Gaia DR3 and an upper limit for the $V$-band extinction from \citep{Schlafly2011}\footnote{\url{https://irsa.ipac.caltech.edu/applications/DUST/}}. A 2.4\% systematic uncertainty floor in \teff\, was adopted, as suggested by \cite{Tayar2022}. 

The SED fitting was performed using the Differential Evolution Markov Chain Monte Carlo (DE-MCMC) technique, integrated within \exofasttwo\, \citep{Eastman2017, Eastman2019} to evaluate uncertainties. The MCMC procedure was considered converged when the Gelman-Rubin diagnostic \citep[$\hat{R}$;][]{Gelman1992} fell below 1.01 and the count of independent draws surpassed 1000. The resulting stellar parameters are listed in Table~\ref{tab:results}.

\begin{figure*}[t]
    \centering
    \includegraphics[width=1\linewidth]{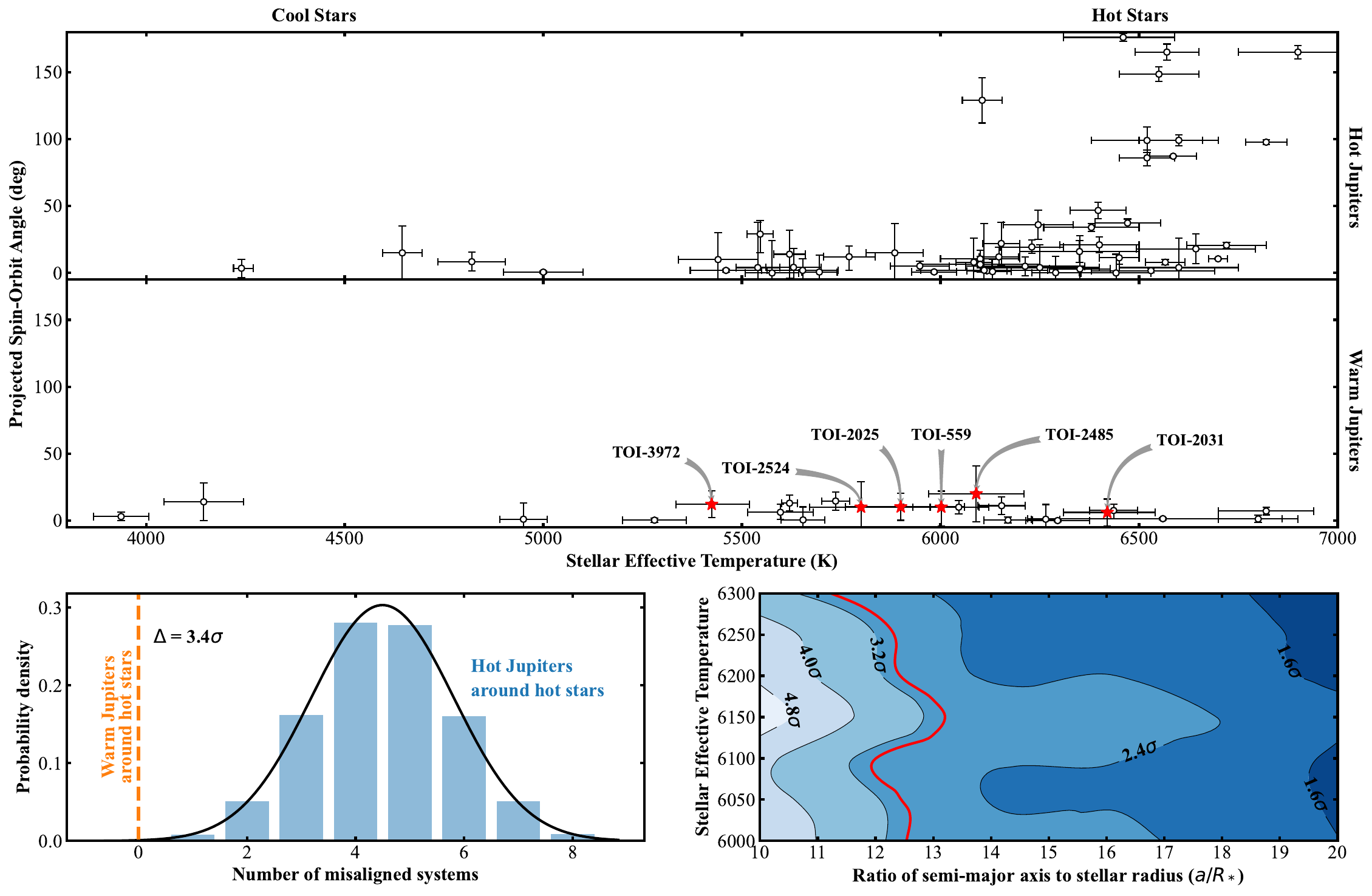}
    \caption{ \textbf{Upper panel}: the distribution of sky-projected stellar obliquity ($\rm{|\lambda|}$) for Jovian planets around hot and cool stars as a function of stellar effective temperature. The obliquity measurements conducted by this work are shown as red stars. \textbf{Lower panel}: \textit{Left}: The greater alignment of hot-star warm-Jupiter systems compared to hot-star hot-Jupiter systems is shown, with the number of misaligned systems for each type represented in orange and blue, respectively. The black line represents the fitted Gaussian function profile. \textit{Right}: significance levels as a function of \teff\, and \ar\,. The brighter regions have higher significance levels. Black lines indicate significance levels, with inline numbers showing the values. The 3$\sigma$ significance level lines are shown in red. The dataset used to create this figure is available via this \href{https://github.com/wangxianyu7/Data_and_code/tree/main/WarmJupitersAreAligned}{link}.}
    \label{fig:teff_lambda}
\end{figure*}

\section{Obliquity Modeling} \label{sec:obliquity_modeling}

To derive the sky-projected spin-orbit angles, a modified version of \texttt{allesfitter} \citep{allesfitter-paper,allesfitter-code} was adopted to conduct simultaneous fits to transits, radial velocities, and Rossiter-McLaughlin measurements. The transit and radial velocity models were integrated using \texttt{PyTransit} \citep{Pytransit} and \texttt{RadVel} \citep{Fulton2018RadVel}, respectively. The RM model, grounded in \cite{Hirano2011}, was implemented in \texttt{tracit} \citep{Hjorth2021, KnudstrupAlbrecht2022}.

The RM, ground-based photometric, and \tess\, data included within our fits are listed in Table~\ref{obslog}. We also incorporated out-of-transit RVs for TOI-559, TOI-2025, TOI-2524, and, TOI-3972. The RVs of TOI-559, as presented by \cite{Ikwut2022}, were obtained using the 1.5m Tillinghast Reflector Echelle Spectrograph (TRES, \citealt{furesz2008design}) on the 1.5m Tillinghast Reflector and CTIO High-Resolution spectrometer (CHIRON, \citealt{Tokovinin2013}) on the CTIO 1.5-meter telescope. The out-of-transit RVs for TOI-2025 include 16 from TRES, 6 from WIYN, and 39 from the Fiber-fed Echelle Spectrograph (FIES; \citealt{telting2014fies, Telting2014}) on the Nordic Optical Telescope (NOT; \citealt{DjupvikAndersen2010}). For TOI-2025, we included an additional RM observation drawn from \cite{Knudstrup2022}, which consists of 16 RVs from FIES from which \cite{Knudstrup2022} previously derived $\lambda=9^{+36}_{-34}$\degrees. For TOI-2524, an additional eight out-of-transit RVs from CHIRON were adopted. For TOI-3972, 13 additional out-of-transit RVs from NEID were also adopted.

We drew initial guesses for the orbital period ($P$), the reference transit mid-time ($T_{0}$), the cosine of the planet's orbital inclination ($\cos{i}$), the radius ratio of the planet to host ($R_{\mathrm{p}}/R_{\star}$), the sum of the stellar and planetary radii divided by the semimajor axis ($(R_{\star}+R_{\mathrm{p}})/a$), and the radial velocity semi-amplitude ($K$) from the values derived from the Exoplanet Follow-up Observing Program (ExoFOP) website\footnote{\url{https://exofop.ipac.caltech.edu/tess/}}. All fitted parameters were allowed to vary and were initialized with uniform priors. For TOI-559, TOI-2025, TOI-2524, and TOI-3972, which each included out-of-transit RVs in their models, the two eccentricity parameters $\sqrt{e} \sin{\omega}$ and $\sqrt{e} \cos{\omega}$ were each initialized with a value of 0. Because no out-of-transit RVs were included within our models of TOI-2031 and TOI-2485, we fixed their eccentricities to zero. The two transformed quadratic limb-darkening coefficients $q_1$ and $q_2$\footnote{The relations between the transformed ($q_1$, $q_2$) and physical ($u_1$, $u_2$) quadratic limb-darkening coefficients are defined by Equations 15 and 16 in \cite{Kipping2013eccq1q2}: $u_{1}=2 \sqrt{q_{1}} q_{2}$ and $u_{2}=\sqrt{q_{1}}\left(1-2 q_{2}\right)$.} for each in-transit dataset (including each photometric band and each instrument used for the RM observation) were initialized with a value of 0.5. Uniform priors ranging from 0 to 1 were applied to $\cosi$. The sky-projected spin-orbit angle $\lambda$ was initialized with a value of $0^{\circ}$ and allowed to vary between $\pm180^{\circ}$. Furthermore, we incorporated Gaussian priors for macro-turbulence ($\zeta$) and micro-turbulence ($\xi$) to reflect stellar surface motion. These priors were from \cite{Doyle2014Vmac} for macro-turbulence and \cite{Jofre2014Vmic} for micro-turbulence, with a standard deviation of 1 km/s applied.

The \tess\ light curves were detrended using Gaussian Processes with a Matern 3/2 kernel, implemented in \texttt{wotan} \citep{Hippke2019}. For each ground-based transit, a third-order polynomial function was employed to model potential trends.  To accommodate the RV offset in the out-of-transit RVs, both a constant baseline and a jitter term were incorporated. For each RM fit, a polynomial was utilized to address short-term overnight instrumental systematics and stellar variability. To determine the optimal degree of the polynomial, we conducted RM-only fits in which only the \vsini, $\lambda$, and polynomial parameters were allowed to vary, with polynomial degrees ranging from 0 to 3. The best fits were obtained through global optimization using differential evolution, implemented in \texttt{PyDE} \citep{Storn1997DifferentialE}. The polynomial yielding the lowest Bayesian Information Criterion was ultimately selected. Furthermore, to account for potential distortions caused by the long exposure times ($>$1000 s) of our RM observations, we performed exposure interpolation during the fitting process to produce 2-minute sampling for each RM fit.

To sample the posterior distributions of all fitted parameters, we employed the affine-invariant Markov Chain Monte Carlo (MCMC) method as implemented in the \texttt{emcee} \citep{emcee} Python package, using 100 walkers. The total number of accepted steps exceeded 200,000, and we discarded the initial 20\% of these steps as burn-in. The chains were deemed to have converged only if all chains were longer than 50 times their autocorrelation lengths. The derived parameters are tabulated in Table~\ref{tab:results}.

For all six targets, we applied the autocorrelation function as implemented in \texttt{SpinSpotter} \citep{Holcomb2022} and the Generalized Lomb-Scargle periodogram \citep{Zechmeister2009} to the \tess\, light curve. This process did not yield a confident identification of the rotation period in any of the six systems, preventing us from establishing limits on their true obliquity.

\section{Statistical Analysis}\label{sec:stat}

{In the past several years, the number of warm Jupiters with Rossiter-McLaughlin measurements has increased significantly (e.g., Kepler-448 b/KOI-12 b, \citealt{Bourrier2015,Johnson2017}; K2-140 b, \citealt{Rice2021K2140}; K2-232 b, \citealt{Wang2021}; Qatar-6 b, \citealt{Rice2023Q6};  TOI-1478 b, \citealt{Rice2022WJs_Aligned}; TOI-1670 c, \citealt{Lubin2023}; TOI-1859 b, \citealt{Dong2023}; TOI-2202 b, \citealt{Rice2023TOI2202}; TOI-677 b, \citealt{Sedaghati2023, Hu2024PFS}; TOI-3362 b, \citealt{Espinoza2023}; 
TOI-4201 b, \citealt{Gan2024};
TOI-4641 b, \citealt{Bieryla2024RM4641}; and WASP-106 b, \citealt{Wright2023, Harre2023, Prinoth2024}). By combining newly collected RM data with literature RMs and following the procedure described in Appendix~\ref{appendix:Sample_Construction}, we obtained a sample of $\Nall$ Jovian-planet systems without known stellar companions. This includes $\Nhj$ hot-Jupiter systems and $\Nwj$ warm-Jupiter systems, providing a valuable opportunity to explore the connection between stellar obliquity and the dynamical histories of warm Jupiters. Notably, to date, SOLES has contributed 14 of these 23 RM measurements to the warm-Jupiter sample.}

\begin{deluxetable*}{lccccccc}                                                                                                                                                                                                  
\tablecaption{Stellar and Planetary Parameters                                    
\label{tab:results}}                                                                                                           
\tabletypesize{\scriptsize}                                                                                                    
\tablehead{\colhead{ }                                                                                                         &\colhead{TOI-559}                               &\colhead{TOI-2025}                                                                              &\colhead{TOI-2031}                                                                &\colhead{TOI-2485}                                       &\colhead{TOI-2524}                                                        &\colhead{TOI-3972}}                                                                                                                                                        
\startdata                                                                                                                     
\multicolumn{5}{l}{\textbf{Stellar Coordinates and Magnitudes:}}\\                                                             
TIC id                                                                                                                         &209459275                                       &394050135                                                                                       &470127886                                                                         &328934463                                                &169249234                                                                  &284206913     \\                                                                                                                                                           
Gaia DR3                                                                                                                       &5057924844082070016                             &2296342614570999552                                                                             &2299101254886404608                                                               &1443530266145475840                                      &3802583419029393152                                                        &427232186629403264     \\                                                                                                                                                  
V                                                                                                                              &$11.091\pm0.007$                                &$11.599\pm0.021$                                                                                &$11.254\pm0.010$                                                                  &$11.935\pm0.026$                                         &$12.800\pm0.103$                                                           &$11.349\pm0.017$ \\                                                                                                                                                        
\hline \\                                                                                                                      
\multicolumn{5}{l}{\textbf{Stellar Parameters:}}\\                                                                             
\multicolumn{6}{l}{{Synthetic spectral fit:}}\\                                                                                
$\teff$                                                                                                                        &$6037.08\pm121.53$                              &$6086.83\pm80.21$                                                                               &$6415.2\pm105.59$                                                                 &$6137.43\pm96.49$                                        &$5866.50\pm102.60$                                                         &$5420.42\pm70.59$\\                                                                                                                                                        
$\feh$                                                                                                                         &$-0.27\pm0.07$                                  &$0.12\pm0.05$                                                                                   &$-0.21\pm0.07$                                                                    &$0.1\pm0.06$                                             &$0.14\pm0.06$                                                              &$0.21\pm0.05$\\                                                                                                                                                            
$\logg$                                                                                                                        &$4.45\pm0.18$                                   &$4.31\pm 0.16$                                                                                  &$4.44\pm0.22$                                                                     &$4.2\pm0.17$                                             &$ 4.52\pm0.13$                                                             &$4.62\pm0.15$\\                                                                                                                                                            
{$v\sin i_*$}                                                                                                                  &{$3.74\pm0.66$}                                 &{$3.98\pm0.60$}                                                                                 &{$4.66\pm0.91$}                                                   &{$4.39\pm0.62$}                                          &{$2.44\pm0.56$}                                                            &{$2.22\pm0.38$}\\                                                                                                                                                          
\\                                                                                                                             
\multicolumn{6}{l}{{SED+MIST fit (adopted):}}\\                                                                                
$M_*$                                                                                                                          &$0.976^{+0.067}_{-0.059}$                       &$1.129^{+0.12}_{-0.088}$                                                                        &$1.103\pm0.073$                                                                   &$1.262^{+0.084}_{-0.12}$                                 &$1.022^{+0.066}_{-0.059}$                                                  &$0.927^{+0.047}_{-0.037}$\\                                                                                                                                                
$R_*$                                                                                                                          &$1.235^{+0.041}_{-0.040}$                       &$1.524^{+0.014}_{-0.015}$                                                                       &$1.238^{+0.040}_{-0.038}$                                                         &$1.682^{+0.066}_{-0.060}$                                &$1.131^{+0.016}_{-0.014}$                                                  &$0.945^{+0.029}_{-0.028}$\\                                                                                                                                                
$\teff$                                                                                                                        &$6002^{+95}_{-94}$                              &$5900^{+160}_{-140}$                                                                            &$6420^{+120}_{-110}$                                                              &$6090\pm120$                                             &$5800\pm130$                                                               &$5424^{+95}_{-90}$\\                                                                                                                                                       
$\feh$                                                                                                                         &$-0.199^{+0.077}_{-0.11}$                       &$0.13\pm0.13$                                                                                   &$-0.197^{+0.064}_{-0.082}$                                                        &$0.094^{+0.058}_{-0.060}$                                &$0.136^{+0.076}_{-0.080}$                                                  &$0.214^{+0.048}_{-0.050}$\\                                                                                                                                                
$\logg$                                                                                                                        &$4.244^{+0.045}_{-0.043}$                       &$4.130^{+0.058}_{-0.065}$                                                                       &$4.295^{+0.040}_{-0.042}$                                                         &$4.085^{+0.042}_{-0.054}$                                &$4.330\pm0.049$                                                            &$4.455^{+0.032}_{-0.031}$\\                                                                                                                                                
Age                                                                                                                            &$8.2^{+3.0}_{-2.7}$                             &$6.4^{+2.9}_{-2.6}$                                                                             &$3.7^{+2.3}_{-1.8}$                                                               &$4.1^{+2.4}_{-1.2}$                                      &$7.3^{+3.4}_{-3.5}$                                                        &$8.4^{+3.4}_{-3.8}$\\                                                                                                                                                      
$\varpi$                                                                                                                       &$4.317\pm0.019$                                 &$2.930^{+0.017}_{-0.016}$                                                                       &$3.619\pm0.016$                                                                   &$2.527\pm0.056$                                          &$2.315^{+0.020}_{-0.021}$                                                  &$6.132\pm0.017$\\                                                                                                                                                          
d                                                                                                                              &$231.62^{+1.0}_{-0.99}$                         &$341.3^{+1.9}_{-2.0}$                                                                           &$276.3^{+1.3}_{-1.2}$                                                             &$395.7^{+9.0}_{-8.5}$                                    &$432.0^{+4.0}_{-3.6}$                                                      &$163.09^{+0.45}_{-0.44}$\\                                                                                                                                                 
\hline \\                                                                                                                      
\multicolumn{5}{l}{\textbf{Rossiter-McLaughlin Parameters:}}\\                                                                 
$\lambda$                                                                                                                      &$10_{-14}^{+12}$                                &$-10.2_{-10.0}^{+6.9}$                                                                           &$-6.1\pm7.5$                                                                      &$-20_{-21}^{+26}$                                        &$10._{-22}^{+19}$                                                          & $-12.1\pm8.5$ \\                                                                                                                                                      
{$v \sin{i_\star}$}                                                                                                            &{ $2.93\pm0.38$ }                               &$5.43_{-0.93}^{+0.99}$                                                                          &{ $6.56\pm0.60$ }                                                                 &{ $4.26_{-0.69}^{+0.96}$ }                               &{ $2.38_{-0.62}^{+0.57}$ }                                                 & { $1.20\pm0.16$ }\\                                                                                                                                                       
$\xi$                                                                                                                          &$1.26_{-0.76}^{+0.93}$                          &$1.29_{-0.77}^{+0.89}$                                                                          &$1.27_{-0.76}^{+0.92}$                                                            &$1.48_{-0.83}^{+0.94}$                                   &$1.27_{-0.77}^{+0.91}$                                                     & $1.18_{-0.73}^{+0.88}$  	\\                                                                                                                                                 
$\zeta$                                                                                                                        &$3.90\pm1.0$                                    &$4.68\pm0.95$                                                                                   &$5.44\pm0.99$                                                                     &$4.68\pm0.99$                                            &$3.40\pm1.0$                                                               & $2.11\pm0.99$ 	\\                                                                                                                                                  
\hline \\                                                                                                                      
\multicolumn{5}{l}{\textbf{Planetary Parameters:}}\\                                                                           
$\frac{R_b}{R_\star}$                                                                                                          &$0.09211_{-0.00095}^{+0.00088}$                 &$0.07604_{-0.0010}^{+0.00098}$                                                                  &$0.10557\pm0.00061$                                                               &$0.0682\pm0.0022$                                        &$0.1016_{-0.0028}^{+0.0017}$                                               & $0.12564_{-0.00049}^{+0.00042}$           \\                                                                                                                        
$\frac{(R_\star + R_b)}{a_b}$                                                                                                  &$0.0933\pm0.0026$                               &$0.1318_{-0.0052}^{+0.0043}$                                                                    &$0.0989\pm0.0015$                                                                 &$0.0782_{-0.0024}^{+0.0037}$                             &$0.0787_{-0.0019}^{+0.0020}$                                               &  $0.05260\pm0.00047$            \\                                                                                                                       
$\cos{i_b}$                                                                                                                    &$0.0340_{-0.0049}^{+0.0042}$                    &$0.172_{-0.028}^{+0.021}$                                                                       &$0.0362_{-0.0038}^{+0.0035}$                                                      &$0.020_{-0.012}^{+0.011}$                                &$0.0333_{-0.011}^{+0.0043}$                                                & $0.03093\pm0.00064$   \\                                                                                                                                
$T_{0;b}$                                                                                                                      &$9312.84866\pm0.00067$                          &$9400.05721\pm0.00066$                                                                          &$9634.56512\pm0.00010$                                                            &$9490.2896\pm0.0029$                                     &$9469.94454\pm0.00050$                                                     & $9356.64809\pm0.00079$             \\                                                                                                                      
$P_b$                                                                                                                          &$6.9839122\pm0.0000070$                         &$8.872079_{-0.000011}^{+0.000010}$                                                              &$5.7154834\pm0.0000020$                                                           &$11.234813\pm0.000061$                                   &$7.1858163\pm0.0000063$                                                    & $10.511372\pm0.000015$           \\                                                                                                                        
$K_b$                                                                                                                          &$0.6374\pm0.0063$                               &$0.375\pm0.018$                                                                                 &$0.057_{-0.029}^{+0.031}$                                                         &$0.205\pm0.029$                                          &$0.0674\pm0.0045$                                                          & $0.521\pm0.025$                                               \\                                                                                    
$\sqrt{e_b} \cos{\omega_b}$                                                                                                    &$0.181\pm0.010$                                 &$-0.039\pm0.077$                                                                                &$0.0$ (fixed)                                                                     &$0.0$ (fixed)                                                   &$-0.02_{-0.15}^{+0.15}$                                                    &$-0.193\pm0.011$                                                                                              \\                                                           
$\sqrt{e_b} \sin{\omega_b}$                                                                                                    &$-0.346\pm0.016$                                &$0.670_{-0.015}^{+0.013}$                                                                       &$0.0$ (fixed)                                                                     &$0.0$ (fixed)                                                   &$-0.137_{-0.066}^{+0.084}$                                                 &$0.4802_{-0.010}^{+0.0078}$                                                                      \\                                                                        
$q_{1,\rm FPS}$ &$0.50\pm0.34$&-&-&-&-&-\\
$q_{2,\rm FPS}$ &$0.42_{-0.27}^{+0.36}$&-&-&-&-&-\\
$q_{1,\rm NEID}$ &-&$0.56_{-0.34}^{+0.30}$&$0.55_{-0.34}^{+0.31}$&$0.57_{-0.35}^{+0.30}$&$0.58_{-0.35}^{+0.29}$&$0.60_{-0.35}^{+0.29}$\\
$q_{2,\rm NEID}$ &-&$0.48_{-0.27}^{+0.31}$&$0.51\pm0.31$&$0.50_{-0.28}^{+0.30}$&$0.53\pm0.30$&$0.52_{-0.29}^{+0.31}$\\
$q_{1,\rm FIES}$ &-&$0.50\pm0.34$&-&-&-&-\\
$q_{2,\rm FIES}$ &-&$0.55_{-0.38}^{+0.31}$&-&-&-&-\\
\hline \\                                                                                                                      
\multicolumn{5}{l}{\textbf{Derived Parameters:}}\\                                                                             
$M_b$                                                                                                                          &$7.13\pm0.53$                                   &$5.46_{-0.41}^{+0.37}$                                                                          &-                                                                                 &-                                                        &$0.648_{-0.060}^{+0.064}$                                                  &$3.67\pm0.23$ 	\\                                                                                                                                                          
$R_b$                                                                                                                          &$1.085\pm0.031$                                 &$1.127\pm0.019$                                                                                 &$1.272\pm0.041$                                                                   &$1.116\pm0.056$                                          &$1.116_{-0.032}^{+0.025}$                                                  & $1.126\pm0.027$ 	\\                                                                                                                                              
$\frac{a_b}{R_\star}$                                                                                                          &$11.71_{-0.31}^{+0.33}$                         &$13.17_{-0.26}^{+0.34}$                                                                          &$11.17\pm0.17$                                                                    &$13.65_{-0.61}^{+0.43}$                                  &$13.99\pm0.35$                                                             & $21.40\pm0.19$     \\                                                                                                                                                
$a_b$                                                                                                                          &$0.0659\pm0.0026$                               &$0.0579_{-0.0019}^{+0.0025}$                                                                    &$0.0644\pm0.0022$                                                                 &$0.1064_{-0.0060}^{+0.0055}$                             &$0.0736\pm0.0021$                                                          & $0.0917\pm0.0023$  	\\                                                                                                                                           
$i_b$                                                                                                                          &$88.05_{-0.24}^{+0.28}$                         &$80.1_{-1.2}^{+1.7}$                                                                            &$87.93_{-0.20}^{+0.22}$                                                           &$88.86_{-0.65}^{+0.70}$                                  &$88.09_{-0.25}^{+0.66}$                                                    & $88.228\pm0.037$    \\                                                                                                                                                
$e_b$                                                                                                                          &$0.1521\pm0.0092$                               &$0.455\pm0.017$                                                                                 &-                                                                                 &-                                                        &$0.036_{-0.023}^{+0.036}$                                                  &$0.2680_{-0.0072}^{+0.0063}$ 	\\                                                                                                                                           
$\omega_b$                                                                                                                     &$297.6_{-2.1}^{+2.2}$                           &$93.3\pm6.6$                                                                                    &-                                                                                 &-                                                        &$259_{-45}^{+59}$                                                          &$112.0_{-1.3}^{+1.4}$ 	\\                                                                                                                                                  
$T_{14;b}$                                                                                                                     &$5.193_{-0.031}^{+0.033}$                       &$3.91_{-0.25}^{+0.31}$                                                                          &$4.029\pm0.013$                                                                   &$6.491_{-0.094}^{+0.11}$                                 &$4.016_{-0.079}^{+0.16}$                                                   & $3.420\pm0.013$  \\                                                                                                                                                        
$T_{23;b}$                                                                                                                     &$4.113_{-0.045}^{+0.049}$                       &$2.67_{-0.46}^{+0.52}$                                                                          &$3.122\pm0.029$                                                                   &$5.58\pm0.12$                                            &$3.06_{-0.12}^{+0.29}$                                                     & $2.146\pm0.021$ 	\\                                                                                                                                 
\enddata                                                                                                                                                                                              
\tablenotetext{Note:}{\\$\teff$: effective temperature (K),
$\feh$: metallicity (dex),
$\logg$: surface gravity (log$_{10}$(cm/s$^2$)),
{$v\sin i_*$}: projected stellar rotational velocity (km/s),
$M_*$: stellar mass ($M_\odot$),
$R_*$: stellar radius ($R_\odot$),
$\teff$: effective temperature (K),
$\feh$: metallicity (dex),
$\logg$: surface gravity (log$_{10}$(cm/s$^2$)),
Age: age (Gyr),
$\varpi$: parallax (mas),
d: distance (pc),
$\lambda$: sky-projected spin-orbit angle (deg),
{$v \sin{i_\star}$}: projected stellar rotational velocity (km/s), 
$\xi$: micro-turbulent velocity (km/s),
$\zeta$: macro-turbulent velocity (km/s),
$R_b / R_\star$: planet-to-star radius ratio,
{$(R_\star + R_b) / a_b$}: ratio of the sum of star and planet radii to semi-major axis,
$\cos{i_b}$: cosine of inclination,
$T_{0;b}$: mid-transit time - 2450000 ($\rm BJD_{\rm TDB}$),
$P_b$: orbital period (days),
$K_b$: radial velocity semi-amplitude (m/s),
$\sqrt{e_b} \cos{\omega_b}$ and $\sqrt{e_b} \sin{\omega_b}$: eccentricity vector components,
$q_1$ and $q_2$: transformed limb-darkening coefficients,
$M_b$: planetary mass ($\mathrm{M_{jup}}$),
$R_b$: planetary radius ($\mathrm{R_{jup}}$),
$a_b / R_\star$: semi-major axis scaled by stellar radius,
$a_b$: semi-major axis (AU),
$i_b$: inclination (deg),
$e_b$: eccentricity,
$\omega_b$: argument of periastron,
$T_{14;b}$: total transit duration (hours),
$T_{23;b}$: full transit duration (hours).}
\end{deluxetable*}   

\subsection{Statistical Significance of the Aligned Single-Star Warm-Jupiter System Trend}
{We subsequently utilized this sample to examine whether warm-Jupiter and hot-Jupiter systems exhibit similar or distinct misalignment distributions. In this work, we define a ``misaligned'' system as one in which the angle $|\lambda|$ exceeds 10\degrees\, and differs from 0\degrees\, at the 3$\sigma$ level \citep{WangX2022WASP148, Rice2022WJs_Aligned}. All other systems are considered ``aligned''.}

Previous studies have shown that hot-Jupiter systems are often misaligned in the presence of host stars with high stellar effective temperatures (above the Kraft break, $\teff \approx6100$ K, \citealt{kraft1967break}). However, due to a previously small sample size, it has so far been unclear whether this host star temperature trend extends to warm-Jupiter systems. The larger, clean sample of hot- and warm-Jupiter systems constructed within this work provides the necessary context for a comparative study examining the dependence of stellar obliquities on host star temperature.

\noindent \textit{\ul{Single-cool-star systems with giants tend to be aligned:}}

{In systems with cool stars (\teff$<$ 6100 K), there are 19 hot-Jupiter systems and 14 warm-Jupiter systems with RM measurements, all of which are aligned. Consequently, to date, no robust observational evidence} suggests that Jupiters around single cool stars can be spin-orbit misaligned.

To compare if the projected stellar obliquity distributions of cool-star hot-Jupiter and cool-star warm-Jupiter systems are identical, we applied the Kolmogorov-Smirnov (K-S, \citealt{Hodges1958TheSP}) and Anderson-Darling (A-D, \citealt{scholz1987k}) tests, implemented in \texttt{scipy} \citep{virtanen2020scipy}. The null hypothesis is that these distributions are the same. We resampled the $|\lambda|$ values with associated uncertainties and conducted 10,000 trials to derive the distribution of the p-values for the K-S and A-D tests. To be conservative, we adopted the larger value between the lower and upper uncertainties for each $|\lambda|$. The resulting distribution of p-values for the K-S and A-D tests showed that merely 4.3\% and 1.6\% of the values, respectively, fell below the 0.05 threshold, failing to reject the null hypothesis. This indicates that the alignment tendency of cool-star hot-Jupiter systems are similar to those of cool-star warm-Jupiter systems. 

\noindent \textit{\ul{Single-hot-star warm-Jupiter systems tend to be aligned}:}

{Our sample includes \Nhjhs\ hot Jupiters and \Nwjhs\ warm Jupiters around hot stars (\teff$\geq$ 6100 K), with all of these warm Jupiters exhibiting spin-orbit alignment.} To determine the misalignment probability in hot-Jupiter systems, we iteratively draw random samples of \Nwjhs\, $|\lambda|$ values from the hot-star hot-Jupiter sample without replacement to see how many of them are misaligned. This process was repeated until the distribution of the number of misaligned systems stabilized, which occurred after 100,000 iterations. {Since all nine warm Jupiters around hot stars are aligned,} we found a \SigHotStar$-\sigma$ difference between the hot-star hot-Jupiter systems and the hot-star warm-Jupiter systems, as shown in the lower left panel of Figure~\ref{fig:teff_lambda}. Furthermore, we examine the dependency of statistical significance on the choice of the difference criteria, varying $\ar$ from 10 to 20 and \teff\, from 6100 to 6300 K (see the lower right panel of Figure~\ref{fig:teff_lambda}). For 98$\%$ of the area where $6000\leq$\teff$\leq6300$ K and $10\leq\ar\leq12$, the significance level is $\geq 3 \sigma$. 
{A more straightforward approach to verify this result is as follows: among \Nhjhs\ hot-star hot-Jupiter systems, \NhjhsAligned\ are misaligned, corresponding to a misalignment rate of 50\%. The probability of randomly drawing nine systems and finding that all are aligned is $0.50^{9} \approx 0.2\%$, indicating a 3.1-$\sigma$ occurrence.}

Additionally, we treat hot-Jupiter and warm-Jupiter systems as two groups without distinguishing between aligned and misaligned. We applied the K-S and A-D tests utilizing the aforementioned approach to study whether the projected stellar obliquity distribution of hot-star hot-Jupiter and hot-star warm-Jupiter systems is identical. The null hypothesis is that they are identical. The analysis of the p-value distributions demonstrated that 70\% of the p-values from the K-S test and 85\% of the p-values from the A-D test fell below the 0.05 threshold, allowing us to reject the null hypothesis and highlighting the significant disparity between the two samples.

\section{Theoretical Implications}\label{sec:Theoretical_Implication}

In our analysis of Rossiter-McLaughlin measurements, we identified a {\SigHotStar}-$\sigma$ trend\footnote{{Two more anticipated RM measurements for hot-star warm-Jupiter systems from \cite{author2024prep} and \cite{Knudstrup2024} will increase the significance level to 3.9$\sigma$.}}: single-star warm-Jupiter systems consistently exhibit no spin-orbit misalignments, even above the Kraft break, demonstrating a {\SigHotStar}-$\sigma$ confidence difference when compared to the \teff-$\lambda$ dependency found in hot-Jupiter systems. This trend presents intriguing contrasts with the orbital orientations of hot Jupiters, which typically align with cooler stars but often show misalignment around hotter ones.

One explanation for these contrasting alignment trends is that hot and warm Jupiters might originate from different formation processes. Warm Jupiters could form directly from well-aligned protoplanetary disks, leading to an initial alignment (albeit with some low-level, $\lesssim20\degrees$ range of expected primordial misalignments; see \citealt{Rice2023TOI2202}). {This interpretation is in agreement with findings from \citet{Morgan2024}, which found that giant planets around cool stars out to $\sim2$ AU are preferentially aligned.} It is also supported by a growing body of evidence for primordial alignment from direct imaging observations of giant planets around young stars \citep{Kraus2020, Bowler2023, Franson2023, Sepulveda2024} and from both protoplanetary \citep{davies2019star} and debris disk \citep{hurt2023evidence} spin-orbit constraints. By contrast, hot Jupiters, irrespective of stellar host temperature, may emerge from more violent evolutionary processes -- often associated with high-eccentricity migration -- after protoplanetary disk dispersal, resulting in initial misalignment. Subsequently, only cooler stars with hot Jupiters undergo tidal realignment, shaping the observed obliquity distribution in hot-Jupiter systems \citep{Albrecht2012, Rice2022, zanazzi2024damping}. 

The observed sharp transitions near the Kraft break in the relationship between a star's effective temperature and the alignment of hot Jupiter orbits strongly support tidal realignment mechanisms. However, these observations do not preclude the possibility that hot Jupiters around both cool and hot stars might initially have different obliquity distributions.  Hot Jupiters around cooler stars, which may have formed and evolved quiescently, tend to be aligned \citep{Wu2023HJsNotAlone, Hixenbaugh2023}. By contrast, hot Jupiters around hot stars may often experience violent formation histories that lead to significant spin-orbit misalignments \citep{Wu2023HJsNotAlone}. 

This scenario may arise if, for example, higher-mass stellar hosts preferentially form multiple Jupiter-mass planets. Following the dispersal of the protoplanetary disk, planet-planet interactions would naturally increase eccentricities and excite spin-orbit misalignments. As these orbits circularize, the initially eccentric and misaligned warm and cold Jupiters evolve into misaligned hot Jupiters around hotter stars. This explanation also accounts for why current observations show warm Jupiters around hot stars as aligned. 

The fundamental factor here is the relative ease of exciting eccentricities over inducing inclinations \citep{Petrovich20152015CHEM, Espinoza2023, Sedaghati2023}, suggesting that warm Jupiters with significant initial misalignments likely underwent high eccentricities that drove their inward migration, transforming them into misaligned hot Jupiters. The persistent misalignment in resulting hot Jupiters, even though their orbital eccentricity has been damped, can primarily be attributed to differences in tidal damping timescales between orbital circularization and spin-orbit realignment. The timescale for circularization, dominated by tides raised by the star on the planet, is typically much shorter than the timescale for realignment, driven by tides the planet induces on the star \citep{Hut1981, OgilvieANDLin2007}. This disparity in timescales between damping eccentricity and realigning the host star effectively maintains the observed spin-orbit misalignment in these tidally circularized resulting hot Jupiters.

\section*{Acknowledgments}
We appreciate the valuable comments from our anonymous reviewer, which have significantly improved this manuscript. We thank Simon Albrecht, Emil Knudstrup, and {Cristobal Petrovich} for insightful discussions.
We acknowledge support from the NASA Exoplanets Research Program NNH23ZDA001N-XRP (Grant No. 80NSSC24K0153). Additionally, M.R. and S.W. acknowledge support from the Heising-Simons Foundation, with M.R. supported by Grant $\#$2023-4478, and S.W. supported by Grant $\#$2023-4050.
This research was supported in part by Lilly Endowment, Inc., through its support for the Indiana University Pervasive Technology Institute.
The work of HPO has been carried out within the framework of the NCCR PlanetS supported by the Swiss National Science Foundation under grants 51NF40$\_$182901 and 51NF40$\_$205606.
DR was supported by NASA under award number NNA16BD14C for NASA Academic Mission Services.
This paper contains data taken with the NEID instrument, which was funded by the NASA-NSF Exoplanet Observational Research (NN-EXPLORE) partnership and built by Pennsylvania State University. NEID is installed on the WIYN telescope, which is operated by the National Optical Astronomy Observatory, and the NEID archive is operated by the NASA Exoplanet Science Institute at the California Institute of Technology. NN-EXPLORE is managed by the Jet Propulsion Laboratory, California Institute of Technology under contract with the National Aeronautics and Space Administration.
Resources supporting this work were provided by the NASA High-End Computing (HEC) Program through the NASA Advanced Supercomputing (NAS) Division at Ames Research Center for the production of the SPOC data products.
Funding for the TESS mission is provided by NASA's Science Mission Directorate. We acknowledge the use of public TESS data from pipelines at the TESS Science Office and at the TESS Science Processing Operations Center. This research has made use of the Exoplanet Follow-up Observation Program website, which is operated by the California Institute of Technology, under contract with the National Aeronautics and Space Administration under the Exoplanet Exploration Program. This paper includes data collected by the TESS mission that are publicly available from the Mikulski Archive for Space Telescopes (MAST). KAC and CNW acknowledge support from the TESS mission via subaward s3449 from MIT.
This work makes use of observations from the LCOGT network. Part of the LCOGT telescope time was granted by NOIRLab through the Mid-Scale Innovations Program (MSIP). MSIP is funded by NSF.
This paper is based on observations made with the Las Cumbres Observatory’s education network telescopes that were upgraded through generous support from the Gordon and Betty Moore Foundation.
This research has made use of the Exoplanet Follow-up Observation Program (ExoFOP; DOI: 10.26134/ExoFOP5) website, which is operated by the California Institute of Technology, under contract with the National Aeronautics and Space Administration under the Exoplanet Exploration Program.

\clearpage
\vspace{5mm}
\facilities{WIYN/NEID, PFS/Magellan, LCOGT}

\software{\texttt{allesfitter} \citep{allesfitter-paper, allesfitter-code}, \texttt{AstroImageJ} \citep{Collins2017}, \texttt{emcee} \citep{emcee}, \texttt{EXOFASTv2} \citep{Eastman2017, Eastman2019}, \texttt{iSpec} \citep{Blanco2014, Blanco2019}, \texttt{matplotlib} \citep{hunter2007matplotlib}, \texttt{numpy} \citep{oliphant2006guide, walt2011numpy, harris2020array}, \texttt{pandas} \citep{mckinney2010data}, \texttt{PyDE} \citep{Storn1997DifferentialE}, \texttt{PyTransit} \citep{Parviainen2015}, \texttt{scipy} \citep{virtanen2020scipy}, \texttt{TAPIR} \citep{Jensen2013}, \texttt{wotan} \citep{Hippke2019}.}

\appendix 

\section{Ground-based Photometry}
\label{appendix:gb_photometry}
To measure the transit times near the epoch of our spectroscopic time series observations, we acquired ground-based time-series photometry of the fields around TOI-2025 and TOI-2031 as part of the \textit{TESS} Follow-up Observing Program \citep[TFOP;][]{collins:2019}\footnote{\url{https://tess.mit.edu/followup}}. We used the {\tt TESS Transit Finder}, which is a customized version of the {\tt Tapir} software package \citep{Jensen:2013}, to schedule our transit observations. 

\subsection{TOI-2025}
We observed a full transit window of TOI-2025 b in Pan-STARRS $z$-short band on UT 2023 June 14 from the Las Cumbres Observatory Global Telescope (LCOGT, \citealt{Brown2013LCOGT}) 1\,m network node at McDonald Observatory near Fort Davis, Texas, United States. The 1\,m telescope is equipped with a $4096\times4096$ SINISTRO camera having an image scale of $0\farcs389$ per pixel, resulting in a $26\arcmin\times26\arcmin$ field of view. The images were calibrated by the standard LCOGT {\tt BANZAI} pipeline \citep{McCully2018}, and differential photometric data were extracted using {\tt AstroImageJ} \citep{Collins2017}. We used circular photometric apertures with radius $4\farcs3$. The target star aperture excluded all of the flux from the nearest known neighbor in the Gaia DR3 catalog (Gaia DR3 2296342644635200512), which is $\sim80\arcsec$ north of TOI-2025. The light curve data are available on the {\tt EXOFOP-TESS} website\footnote{\url{https://exofop.ipac.caltech.edu/tess/target.php?id=394050135}} and are included in the global modeling described in Section~\ref{sec:obliquity_modeling}. 

\subsection{TOI-2031}
We observed a full transit window of TOI-2031.01 in Sloan $i'$ band on UT 2023 July 06 from the LCOGT 1\,m network node at McDonald Observatory. The images were calibrated by the standard LCOGT {\tt BANZAI} pipeline, and differential photometric data were extracted using {\tt AstroImageJ}. We used circular photometric apertures with radius $6\farcs6$. The target star aperture excluded all of the flux from the nearest known neighbor in the Gaia DR3 catalog (Gaia DR3 2296342644635200512), which is $\sim45\arcsec$ southwest of TOI-2031. The light curve data are available on the {\tt EXOFOP-TESS} website\footnote{\url{https://exofop.ipac.caltech.edu/tess/target.php?id=470127886}} and are included in the global modeling described in Section~\ref{sec:obliquity_modeling}.

\section{Sample Construction}
\label{appendix:Sample_Construction}
We began with the set of systems listed in the TEPCat orbital obliquity catalog\footnote{\url{https://www.astro.keele.ac.uk/jkt/tepcat/obliquity.html}} \citep{Southworth2011} as of July 25, 2024. We adopted the preferred values for $\lambda$ and \teff\, from the TEPCat catalog for each system. Other parameters were drawn from the Planetary Systems Composite Data table of the NASA Exoplanet Archive\footnote{\url{https://exoplanetarchive.ipac.caltech.edu/cgi-bin/TblView/nph-tblView?app=ExoTbls&config=PSCompPars}}. Note that the parameters for WASP-109 and WASP-111 were from \cite{Anderson2014arXiv}. We followed the steps outlined below to define our sample.

\begin{enumerate}

\item \textit{Minimize biases introduced by nonuniform measurement techniques.}
We only included results derived from Rossiter-McLaughlin and Doppler tomography measurements to avoid the biases caused by other methods. For example, spot-crossing is particularly sensitive to systems that are either aligned or anti-aligned, whereas gravity darkening is most sensitive to misaligned systems (\citealt{Siegel2023, Dong2023StellarObliquity}; see \citealt{Albrecht2022} for a review of stellar obliquity measurement methods). When selecting stellar obliquity measurements, we prioritize the most recent results, with Rossiter-McLaughlin measurements receiving higher precedence than Doppler tomography measurements.

\item \textit{Remove low-quality and contested measurements.}
We removed the systems with low-quality or contested measurements as suggested by \cite{Albrecht2022}, including  CoRoT-1 \citep{Bouchy2008, Pont2010}, CoRoT-19 \citep{Guenther2012}, HATS-14 \citep{Zhou2015}, HAT-P-27 \citep{Brown2012}, WASP-1 \citep{Simpson2011a, Albrecht2011}, WASP-2 \citep{Triaud2010, Albrecht2011},  WASP-23 \citep{Triaud2011}, WASP-49 \citep{Wyttenbach2017},  and WASP-134 \citep{Anderson2018a}. Note that we also excluded HAT-P-17 considering the disagreement between its two measurements (\citealt{Fulton2013}, $19^{+14}_{-16}$ \degrees; \citealt{Mancini2022}, $-27.5\pm6.7$\degrees)

\item \textit{Remove binary and multi-star systems.}
Apart from the relatively confined gravitational interactions within a single-star system, stellar obliquity can also be strongly influenced by stellar companions (e.g., \citealt{Wu2003, Fabrycky2007, Naoz2012, Naoz2016}). Therefore, in this work, we have excluded stars with bound companions identified in \textit{Gaia} DR3. Following the methods described in \cite{Badry2021} and \cite{Rice2022WJs_Aligned}, we checked for bound companions to each system, determining whether candidate companions meet the following criteria: 1) their proper motion is within 5 km/s of the planet-hosting star, and 2) their parallax is consistent with that of the planet-hosting star within 5$\sigma$. Candidates satisfying both criteria were identified as stellar companions and subsequently removed from our sample. Moreover, systems listed in the Binary and Multiple Star systems catalog\footnote{\url{https://adg.univie.ac.at/schwarz/intro.html}}\citep{schwarz2016new}, the catalog of Exoplanets in Visual Binaries \citep{Fontanive2021}, or flagged as having more than one star in  NASA Exoplanet Archive were also excluded.

\item \textit{Remove systems with the \teff\, of host stars $<$ 3500 K or $>$ 7000 K.} To ensure uniformity in the effective temperature (\teff) range across both warm- and hot-Jupiter systems, exclusion criteria were applied where \teff\, exceeds 7000 K or falls below 3500 K.

\end{enumerate}

This procedure yielded \Nall\, Jovian planets ($M_{\rm pl} \ge 0.3 \mj$) around single-star systems with stellar obliquity measurements, including \Nhj\, hot-Jupiter systems ($\ar < 11 $) and \Nwj\, warm-Jupiter systems ($\ar \ge 11$). A 0.1 $\mj$ uncertainty was assumed for planetary measurements with only upper-limit values.

\bibliography{main}{}
\bibliographystyle{aasjournal}

\end{document}